\documentclass{webofc}
\usepackage[varg]{txfonts}   
\usepackage{lineno}

\begin{document}
\title{Connection between soft and hard probes of small collision systems at RHIC and LHC}
%
%

\author{\firstname{Dennis V.} \lastname{Perepelitsa}\inst{1}\fnsep\thanks{\email{dvp@colorado.edu}}}

\institute{University of Colorado Boulder, 2000 Colorado Ave, Boulder, CO 80309 USA}

\abstract{%

The expected connections between signatures in the soft and hard
sectors of small collision systems, and the status of experimental
attempts to identify them, are discussed. These proceedings summarize
the talk as given at the International Symposium on Multi-Particle
Dynamics in September 2019 in Santa Fe, NM (ISMD19). As such, the
choice of content and focus are selective and not intended to be comprehensive.}
\maketitle

\section{Introduction}

A significant open challenge in the study of QCD systems and their
interactions is to understand the nature of the system formed in
high-multiplicity proton-proton ($pp$) or proton-nucleus ($p$+A)
collisions~\cite{Nagle:2018nvi,Adolfsson:2020dhm} at the Relativistic
Heavy Ion Collider (RHIC) and the Large Hadron Collider (LHC).

Analyses of the correlations between final-state particles in these
systems reveal that they behave collectively, with their yields at
fixed transverse momentum ($p_\mathrm{T}$) following common,
preferential orientations in the transverse plane. These patterns are
popularly represented as a series of $n$-th order azimuthal anisotropy
coefficients, $v_n$. In nucleus-nucleus (A+A) collisions, this bulk
particle behavior can be well described using theoretical frameworks
based on a viscous hydrodynamical description of final-state
interactions in the Quark-Gluon Plasma (QGP) medium. Remarkably, these
models are also able to describe the $v_n$ values as a function of
$p_\mathrm{T}$ across $pp$ and $p$+A collisions, in a common paradigm
with A+A collisions~\cite{Weller:2017tsr}.

\begin{figure}[t]
\centering
\includegraphics[width=0.46\linewidth]{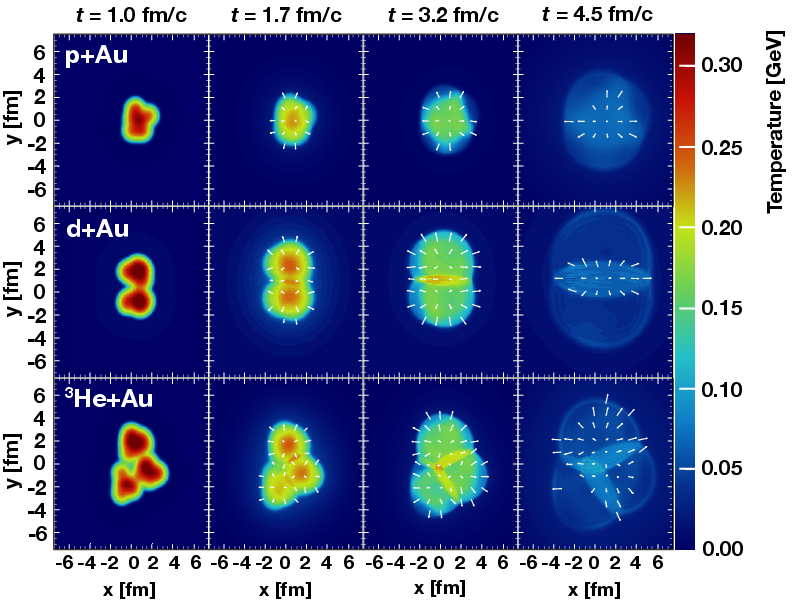}
\includegraphics[width=0.53\linewidth]{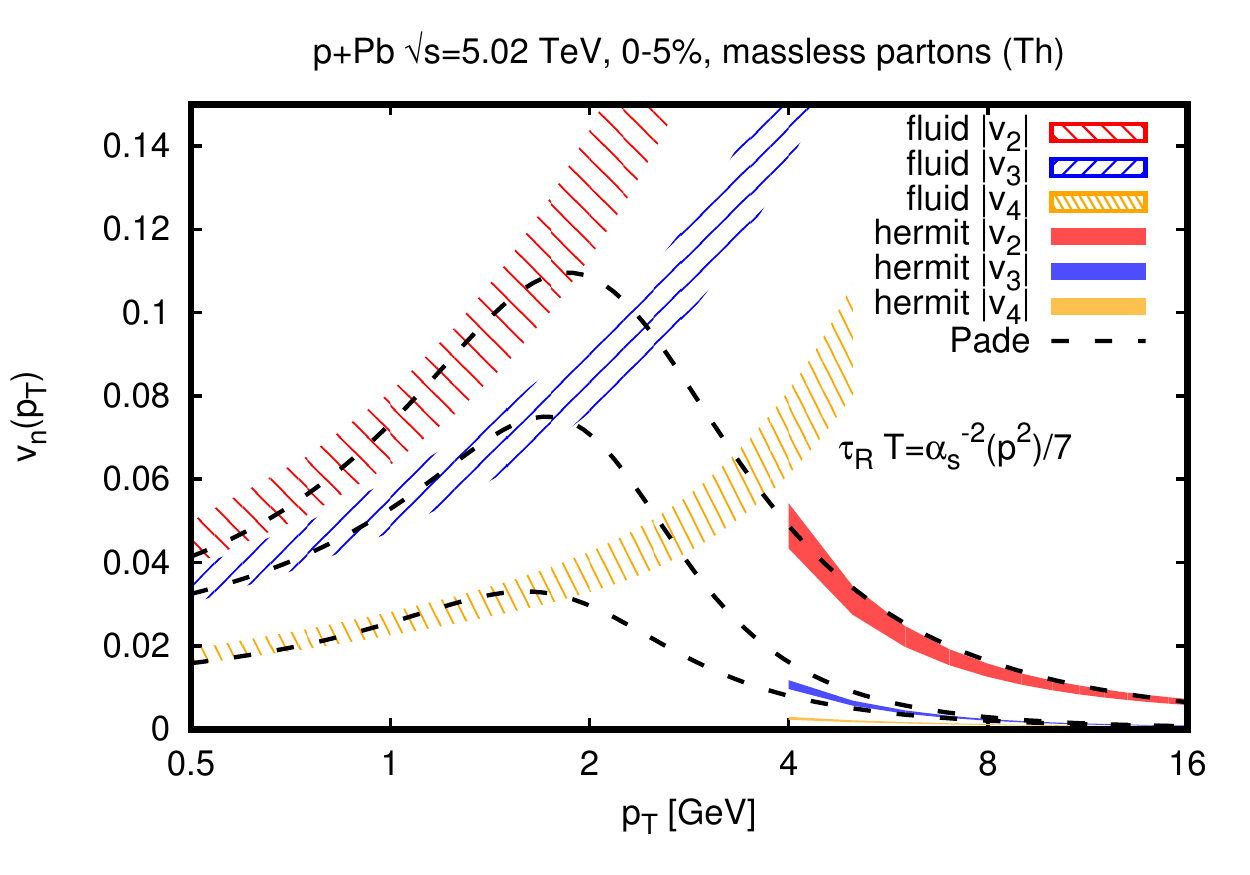}
\caption{{\em Left:} Demonstration of hydrodynamic expansion of QGP
  medium in small collision systems, as part of the RHIC ``projectile
  scan,'' from Ref.~\cite{PHENIX:2018lia}. {\em Right:} Calculated
  $v_{2,3,4}$ values in central $p$+Pb collisions at low and high
  $p_\mathrm{T}$, from Ref.~\cite{Romatschke:2018wgi}.
\label{fig1}}
\end{figure}

This picture is strengthened by a large body of corroborating data,
including but not limited to: the ``projectile scan'' program at
RHIC~\cite{PHENIX:2018lia}, in which $p$+Au, $d$+Au and $^{3}$He+Au
collisions are used to produce systems with differing initial-state
geometric anisotropies $\varepsilon_{2,3}$ which are then observed to be
translated into corresponding differences in the measured $v_{2,3}$
(shown in Fig.~\ref{fig1}); the observation of hadron mass ordering in the
$v_n$ coefficients~\cite{Sirunyan:2018toe}, as expected from motion in
a common velocity field; and measurements of multi-particle
correlations which verify that the observed anisotropies are truly
event-wide phenomena affecting all particles~\cite{Aidala:2017ajz}.

\section{Expectations for the hard sector}

For measurements of particles with $p_\mathrm{T} \lesssim 3$~GeV (the
so-called ``soft sector''), the paradigm described above with
final-state interactions between strongly-interacting particles is
dominant and widely supported by data. The purpose of these
proceedings is to discuss what, given this paradigm, may be expected
in the so-called ``hard sector'', where particle production is
dominated by the fragmentation of high-$p_\mathrm{T}$ partons that
emerge as jets.  Rather than being treated as two unrelated phenomena,
one should strive to understand the two behaviors together, since
ultimately they are taking place concurrently in the same physical
system.

In A+A collisions, the dominant behaviors in soft and hard sectors are
indirectly connected through the common relationship of each to the
underlying geometry of the collision: low-$p_\mathrm{T}$ particles
acquire an azimuthal anisotropy from the larger pressure gradient
along the minor axis of the elliptic overlap region during the
hydrodynamic expansion, while high-$p_\mathrm{T}$ particles acquire it
from azimuthally-differential energy loss. Thus
both low-$p_\mathrm{T}$ and high-$p_\mathrm{T}$ particle pair yields follow a
$\cos(2\Delta\phi)$ modulation. In fact, attempting to simultaneously
understand differential energy loss and hydrodynamic flow in A+A collisions
has spurred important theoretical
developments~\cite{Noronha-Hostler:2016eow}.

One direct calculation of what may be expected at high-$p_\mathrm{T}$
in $p$+A collisions is performed in Ref.~\cite{Romatschke:2018wgi},
and is summarized in Fig.~\ref{fig1}. The calculation uses a common
hydrodynamically evolving background and determines the $v_{2,3,4}$
for particles in many-scattering and few-scattering limits at low and
high $p_\mathrm{T}$, respectively. This calculation indicates that
appreciable $v_{2}$ values on the order of $2$\% for $p_\mathrm{T} >
10$~GeV particles should inescapably accompany a non-trivial $v_2$ at
low $p_\mathrm{T}$. However, in most models~\cite{Zhang:2013oca} this
necessarily results in a significant effect in the nuclear
modification factor, $R_{p\mathrm{A}}$, which is not borne out by
data: a large set of jet and charged-particle measurements have
indicated that in minimum-bias $p$+A collisions, there is no
detectable modification.

\begin{figure}[t]
\centering
\includegraphics[width=0.46\linewidth]{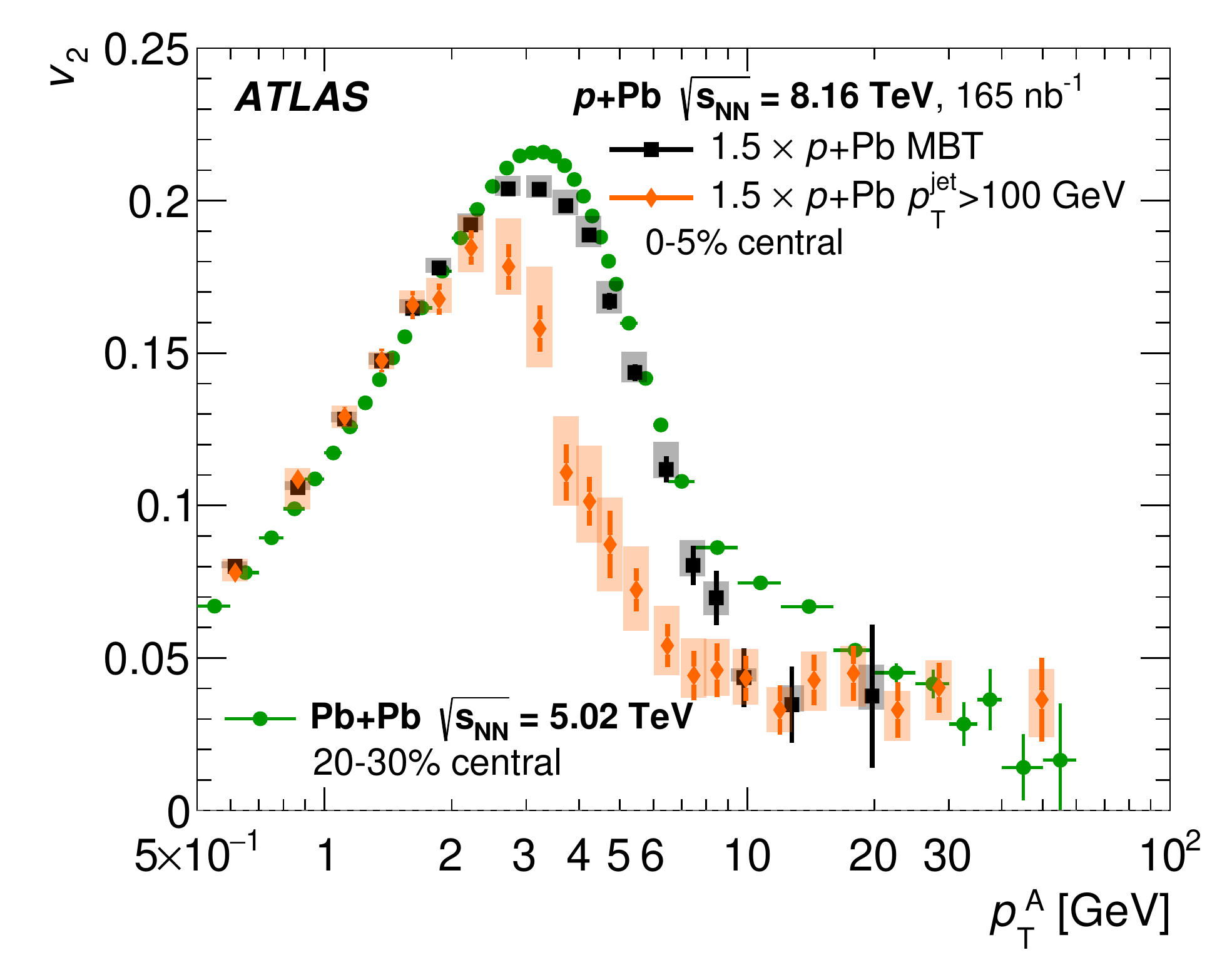}
\includegraphics[width=0.52\linewidth]{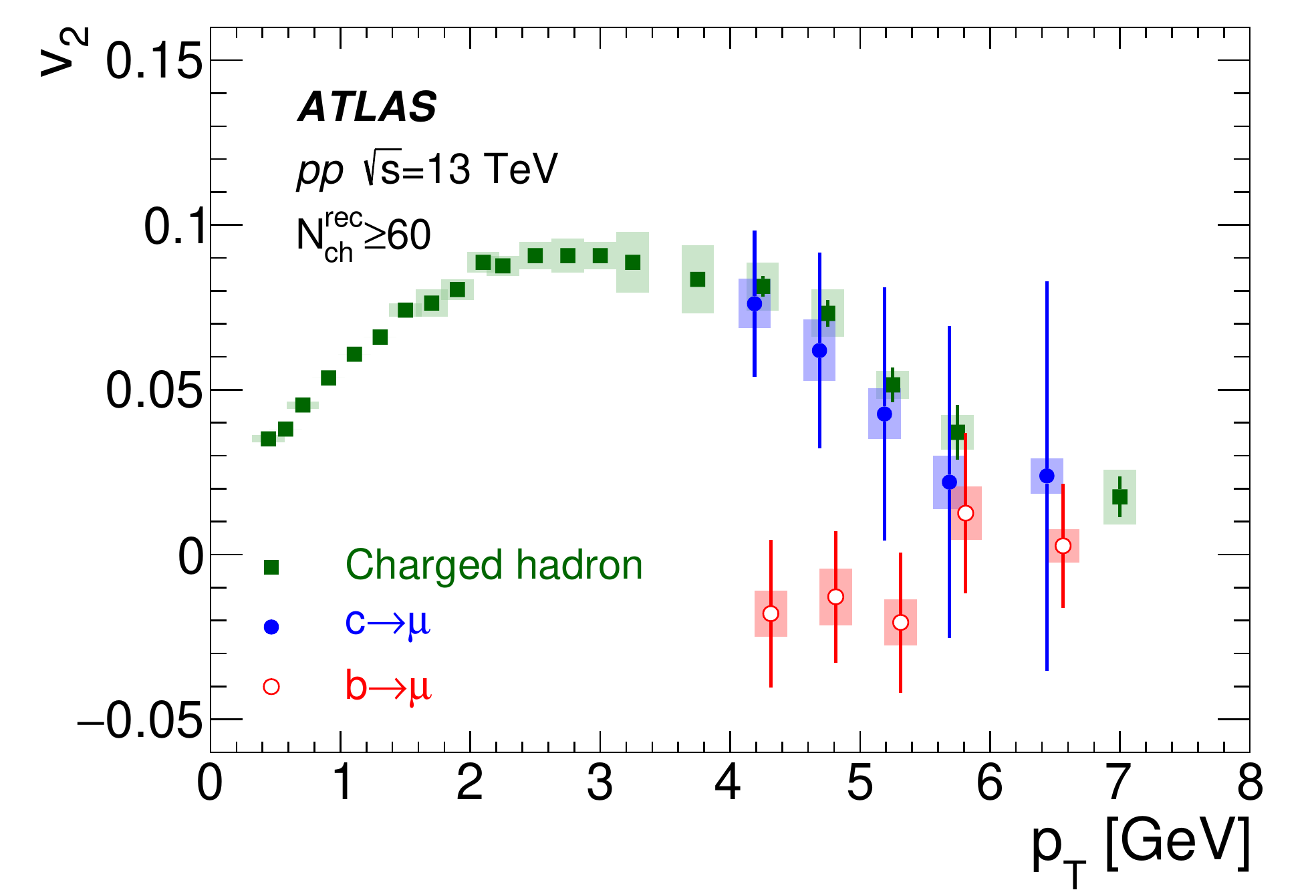}
\caption{{\em Left:} Measured azimuthal anisotropy coefficients $v_2$
  in $p$+Pb collisions using 8.16~TeV data, from
  Ref.~\cite{Aad:2019ajj}. {\em Right:} Measured $v_2$ for muons from
  charm and bottom hadron decays, compared to that for light hadrons,
  in high-multiplicity 13~TeV $pp$ collisions, from
  Ref.~\cite{Aad:2019aol}.
\label{fig2}}
\end{figure}

In the first $p$+A data taken in 2013 at the LHC, the total event
statistics prohibited a detailed study of this high $p_\mathrm{T}$
region, even in ultra high-multiplicity events enhanced by special
triggers. Nevertheless, the early results in Ref.~\cite{Aad:2014lta}
suggested that $v_2$ could have an appreciable value even at
$p_\mathrm{T} \approx 10$~GeV in 0-1\% centrality 5.02~TeV $p$+Pb
events. Shortly after ISMD19, new results from ATLAS were presented in
which jet-triggered events were used in the significantly
higher-luminosity $8.16$~TeV $p$+Pb data taken in
2016~\cite{Aad:2019ajj}. In these results, shown in Fig.~\ref{fig2}, a
non-zero azimuthal anisotropy was observed for particles up to
$p_\mathrm{T} \approx 50$~GeV, far from the hydrodynamic regime, and
over a broad centrality range. The overall shape of the $v_{2}$ as a
function of $p_\mathrm{T}$ is remarkably similar to that in A+A
collisions, again suggesting that A+A-like effects are present in
$p$+A collisions. These results exacerbate the already significant
tension with a minimum-bias $R_{p\mathrm{A}}$ factor of unity.

\subsection{Jet rates and modification}

Of course, azimuthal anisotropies at high $p_\mathrm{T}$ are not the
only possible manifestation of A+A-style partonic energy
loss. Following the analogy with A+A collisions, it is useful to have
measurements of the nuclear modification factor for various
strongly-interacting probes, particularly for system sizes in the
transition region from peripheral A+A to central $p$+A, over which
energy loss effects are rapidly turning off. Unfortunately, this
region is bracketed by the opposite extrema of each type of system,
and each is potentially affected by large biases. Central $p$+A
collisions feature auto-correlations between hard processes and soft
particle production~\cite{Adam:2014qja,Perepelitsa:2014yta}, while
selections on very peripheral A+A collisions may suffer from opposite
(i.e. jet veto) effects~\cite{Morsch:2017brb}. Model-dependent
corrections can be applied to account for these, but it is difficult
to obtain the needed level of precision to observe or rule out energy
loss effects.

Coincidence measurements where jets are paired with an electroweak
boson, another jet, or a high-$p_\mathrm{T}$ trigger particle can
avoid this specific limitation. These measurements are sensitive to
energy loss effects which distort the expected kinematic balance
between final-state objects, and can be reported as a function of
event activity. For example, a measurement by ALICE of semi-inclusive
recoil jet distributions in 5.02~TeV data~\cite{Acharya:2017okq}
provides constraints on the out-of-cone energy loss. However, this
strategy also comes with potential pitfalls: they may have a reduced
sensitivity to energy loss (e.g. a di-jet measurement where both jets
lose energy, leaving them balanced on average) and the multiplicity
selection itself may incur significant biases (e.g. in $pp$
collisions, the resulting enhancement of multi-jet event topologies
may significantly modify particular observables). Nevertheless,
additional coincidence measurements enabled by the high-statistics
8.16~TeV $p$+Pb data would be very valuable.

Finally, the jet-medium interaction in $p$+A collisions may result
only in modest in-cone modification of the parton shower, and thus
give a null result in the above approaches. In this case, precision
measurements of jet structure in $p$+A collisions are needed. The
techniques learned from successful measurements A+A
collisions~\cite{Acharya:2017goa,Aaboud:2019oac}, along with
comparisons to state-of-the-art pQCD calculations such as
SCET$_G$~\cite{Kang:2014xsa}, can be applied here. Some initial
measurements in this direction of fragmentation
functions~\cite{Aaboud:2017tke} and jet masses~\cite{Acharya:2017goa}
have been performed but only in minimum-bias $p$+A collisions where
jet modification effects are likely to be very modest.

\subsection{Heavy flavor probes}

Since high-$p_\mathrm{T}$ light hadrons are produced as part of the
jet fragmentation process, it is possible they are not sensitive to
the QGP medium in $p$+A collisions. This may occur if there is a
finite formation time, or a virtuality separation between the
developing shower of the hard-scattered parton and the short-lived
medium. This possibility can be avoided with heavy flavor quarks,
which due to their large mass must be created early in the collision
and thus are present as a conserved flavor charge throughout the full
evolution of the medium. Historically, the first measurements of
simultaneous flow and suppression of heavy quarks in A+A
collisions~\cite{Adare:2006nq} were important in indicating that a
strongly-coupled medium was indeed being formed at RHIC. Studies at
the LHC indicate that charm quarks approach equilibrium with the QGP
and are thus sensitive to the shape of the produced
region~\cite{Acharya:2018bxo}.

Current data on heavy flavor quark production in $p$+A collisions
indicates no modification at the $10$\% level in production spectra
compared to $pp$ collisions~\cite{Adam:2016ich}, except perhaps at
very low $p_\mathrm{T}$. However, identified $D$-mesons show a
significant azimuthal anisotropy in high-multiplicity events over a
wide $p_\mathrm{T}$ range~\cite{Sirunyan:2018toe}. The combination of
no modification in the $p_\mathrm{T}$ distribution with an appreciable
modulation results in tension: in a final-state-effect picture, how
can the trajectories of the heavy quarks be redirected to such a large
degree without a corresponding modification to the $p_\mathrm{T}$
distribution?

New ATLAS data on the azimuthal anisotropies of heavy flavor muons
bend this contradiction to its breaking point~\cite{Aad:2019aol}, and
are summarized in Fig.~\ref{fig2}. Whereas charm hadrons are observed
to flow with a similar magnitude as light hadrons for a broad range of
$pp$ multiplicities, the anisotropy for bottom hadrons is compatible
with zero! This is one of the rare strongly interacting probes which
is observed to {\em not} participate in collective motion~\footnote{A
  similarly interesting example is the recent indication that the
  $v_2$ for Upsilon particles - a $b\bar{b}$ bound state - is
  compatible with zero in Pb+Pb
  collisions~\cite{Acharya:2019hlv}. Perhaps the common heavy quark
  content is not coincidental.}.  An alternative possibility is that
this signal arises from initial-state momentum
correlations~\cite{Zhang:2019dth}, however this picture has challenges
of its own in describing the multiplicity dependence of the data and
will not be discussed further here.

These data indicate the special role of heavy flavor quarks in
delineating boundaries for final-state collectivity, such as the
degree of thermalization or the system size and lifetime. To determine
the limiting conditions for $b$-quark flow to appear, a precise
measurement in $p$+A collisions would be very valuable. Shortly after
ISMD19, a measurement of non-prompt $D^0$ flow was made by
CMS~\cite{CMS:2019isc}, but within the reported uncertainties it is
compatible with the level of charm flow, or no flow at all.

\section{Novel collision systems}

\begin{figure}[t]
\centering
\includegraphics[width=0.53\linewidth]{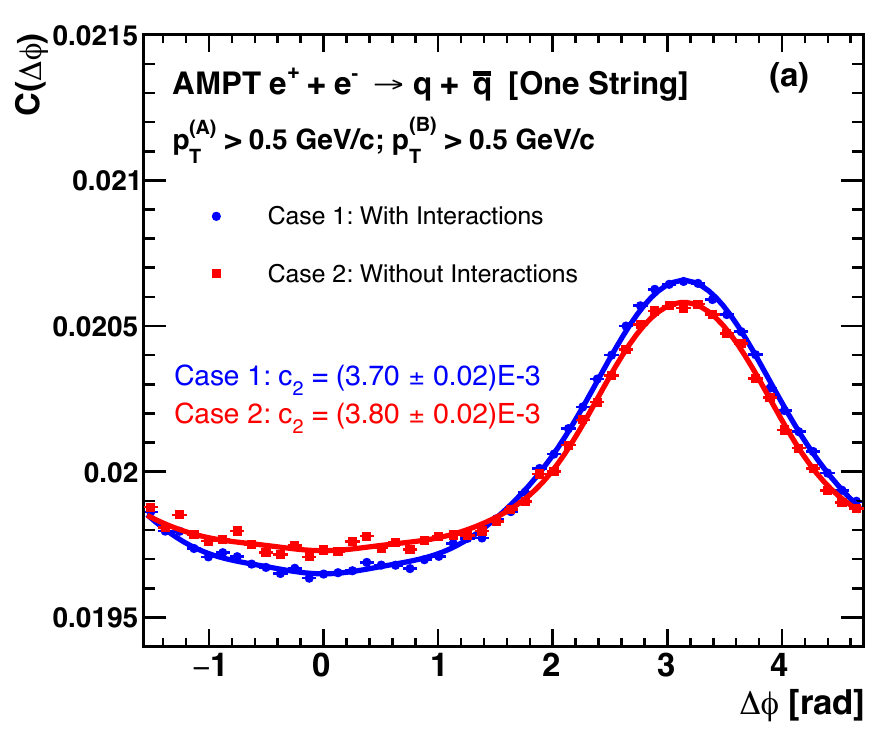}
\includegraphics[width=0.46\linewidth]{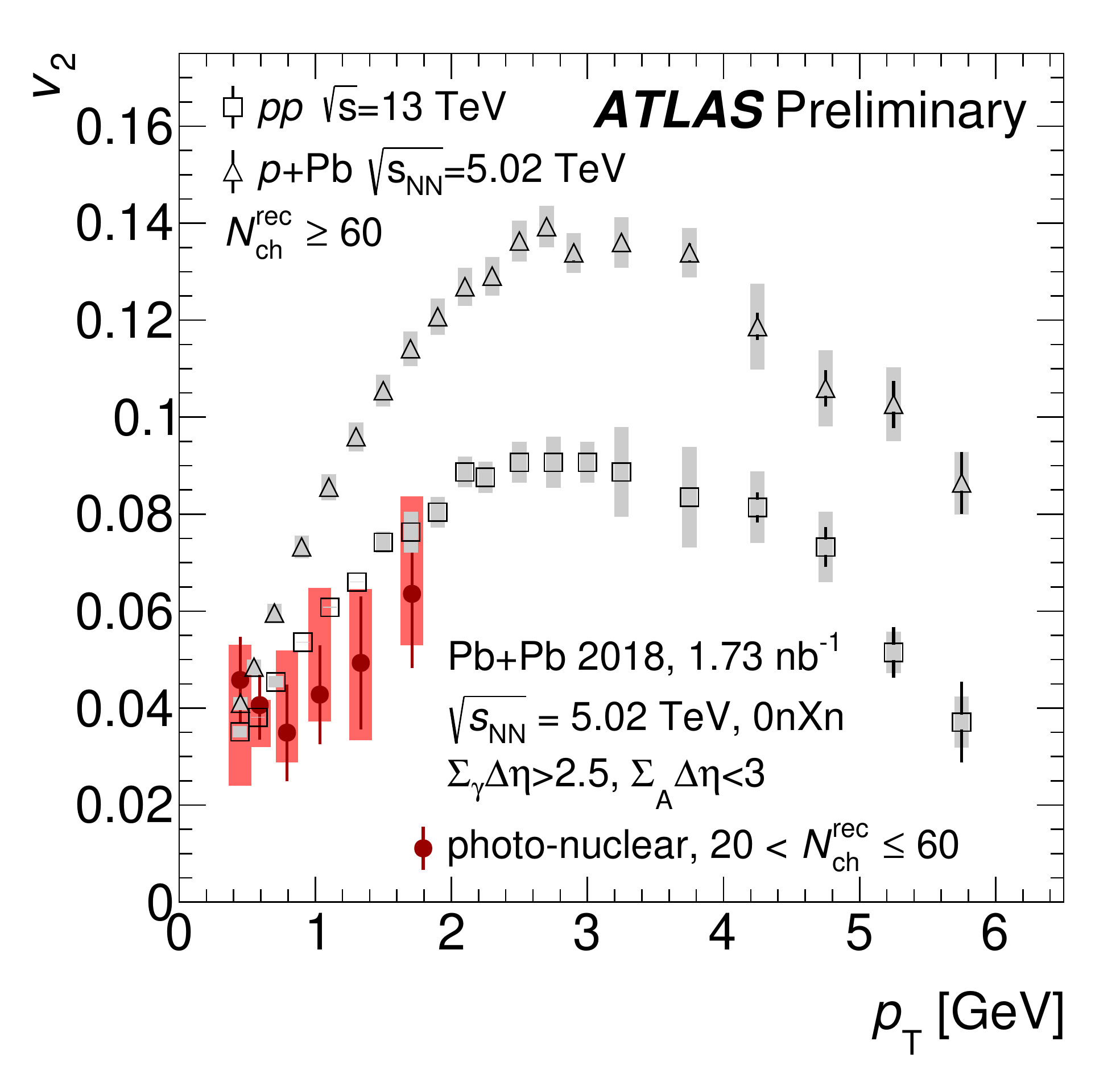}
\caption{{\em Left:} Two-particle $\Delta\phi$ distributed in an AMPT
  simulation of $e^+e^-\rightarrow{q}\bar{q}$ events with a single
  string between the outgoing quarks, from
  Ref.~\cite{Nagle:2017sjv}. {\em Right:} Measured $v_2$ in
  photo-nuclear events, compared to that in $pp$ and $p$+A events,
  from Ref.~\cite{ATLAS-CONF-2019-022}.
\label{fig3}}
\end{figure}

If a deconfined QGP medium is indeed created in $p$+A and $pp$
collisions, one may expect several possible experimental
signatures. For the case of azimuthal anisotropies, the signature
primarily focused on here, a sufficiently high multiplicity is not by
itself determinative. In the final state interaction picture, it is
the underlying transverse spatial geometry of the system which is
translated to a momentum anisotropy in the transverse plane. To better
test this picture, it is interesting to study exotic collision systems
where the transverse geometry is very different and any possible
initial-state effects~\cite{Schenke:2019pmk} have a different relative
contribution. For example, consider the AMPT study described in
Ref.~\cite{Nagle:2017sjv}, which is shown in Fig.~\ref{fig3}. If one
models a $e^+e^- \rightarrow q\bar{q}$ event as a single string
stretched between the outgoing quarks, even with final-state
interactions, no $v_2$ is generated since there is no common
event-wide geometry. On other hand, the fictitious scenario in which
the total energy or string tension is split between two nearby,
parallel strings does generate a $v_2$.

These kinds of pictures can be tested in data: there have been
tremendous efforts to recover old collider data and re-analyze them in
the modern context of two- and multi-particle correlation analyses
which characterize azimuthally anisotropic particle production. The
publically available results so far are a re-analysis of 91~GeV
$e^+e^-$ collisions with ALEPH at LEP~\cite{Badea:2019vey} and 316~GeV
$ep$ collisions with ZEUS at HERA~\cite{ZEUS:2019jya}, with others
underway. In the first measurement, upper statistical limits are set
on the possible magnitude of $v_2$, while the latter does not observe
compatible patterns in the data. Since these experiments and
facilities are no longer operating, it is difficult to imagine
significant improvements in the archived data, and insight may come
only from future data on novel collision systems at RHIC, the LHC, and
eventually the Electron Ion Collider (EIC).

Another interesting possibility is to explore the so-called
ultra-peripheral A+A processes, electromagnetic interactions in which
the nuclear beams remain separated. Among these, photo-nuclear
($\gamma$+A) interactions are those in which one nucleus emits a
quasi-real photon which strikes the other nucleus. These events
feature a significantly cleaner environment for exploring nuclear
effects than $p$+A collisions and allow for the possibility of
accessing ``nuclear-DIS-like'' physics (at least, at the
photoproduction limit of DIS) before the EIC is operating.

These events can also be used to search for collective phenomena, such
as the analysis of high-multiplicity photo-nuclear events in
Ref.~\cite{ATLAS-CONF-2019-022}, shown in Fig.~\ref{fig3}. Using
standard two-particle correlation methods, a symmetric ridge can be
observed in these events. When interpreted as a single-particle
anisotropy under the assumptions of factorization and the non-flow
subtraction procedure, it results in $v_2$ values compatible with that
in $pp$ collisions but lower than that in $p$+Pb. However, this may
not be surprising in a final-state picture: in the vector-meson
dominance (VMD) paradigm, the photon may fluctuate to a meson,
especially under a high multiplicity selection. Thus many of these
collisions may proceed as $\rho$+A or $\omega$+A collisions with a
non-trivial transverse geometric structure which may induce
momentum-space anisotropies. A more quantitative comparison to theory
which matches the particular $\gamma$+A kinematics and experimental
acceptance would be very interesting.

\section{Conclusion}


To better understand the connection between soft and hard probes of
small collision systems, several specific approaches can be
proposed. Precision jet-coincidence and jet-structure measurements
should continue to be performed in $p$+A collisions, ideally with
guidance from theoretical approaches that correctly describe A+A
collisions. Measurements of azimuthal anisotropy should be extended to
higher $p_\mathrm{T}$ and should particularly be performed for charm
and bottom quarks as key discriminants. New collision systems at RHIC
and the LHC, such as possible O+O or Ar+Ar
running~\cite{Citron:2018lsq}, are important for filling in the $p$+A
/ A+A ``gap'' without biases arising from extreme multiplicity
selections. Finally, novel systems such as those accessible in
ultra-peripheral collisions and at the future EIC will be important
for testing the final-state interaction picture in situations with
starkly different geometries.

%
\bibliography{dvp-ISMD2019-proceedings}
%
%
%
%

\end{document}